\documentclass[sigconf]{acmart}

\AtBeginDocument{%
  \providecommand\BibTeX{{%
    \normalfont B\kern-0.5em{\scshape i\kern-0.25em b}\kern-0.8em\TeX}}}

\copyrightyear{2024}
\acmYear{2024}
\setcopyright{acmlicensed}\acmConference[DIS Companion '24]{Designing Interactive Systems Conference}{July 1--5, 2024}{IT University of Copenhagen, Denmark}
\acmBooktitle{Designing Interactive Systems Conference (DIS Companion '24), July 1--5, 2024, IT University of Copenhagen, Denmark}
\acmDOI{10.1145/3656156.3663694}
\acmISBN{979-8-4007-0632-5/24/07}

\usepackage{multirow}
\usepackage{subcaption}

\graphicspath{{images/}}
\acmSubmissionID{pwip3762}

\begin{document}

\title{Towards Understanding Emotions for Engaged Mental Health Conversations}

\author{Kellie Yu Hui Sim}
\affiliation{%
  \institution{Singapore University of Technology and Design}
  \country{Singapore}
  \city{Singapore}}
\email{kelliesyhh@gmailcom}
\orcid{0009-0005-6451-7089}

\author{Kohleen Tijing Fortuno}
\affiliation{%
  \institution{Singapore University of Technology and Design}
  \country{Singapore}
  \city{Singapore}}
\email{kohleen.tf@gmail.com}
\orcid{0000-0002-3789-6926}

\author{Kenny Tsu Wei Choo}
\affiliation{%
  \institution{Singapore University of Technology and Design}
  \country{Singapore}
  \city{Singapore}}
\email{kennytwchoo@gmail.com}
\orcid{0000-0003-3845-9143}

\renewcommand{\shortauthors}{Sim, Fortuno, and Choo}

\begin{abstract}
    Providing timely support and intervention is crucial in mental health settings.
As the need to engage youth comfortable with texting increases, mental health providers are exploring and adopting text-based media such as chatbots, community-based forums, online therapies with licensed professionals, and helplines operated by trained responders. 
To support these text-based media for mental health--particularly for crisis care--we are developing a system to perform passive emotion-sensing using a combination of keystroke dynamics and sentiment analysis. 
Our early studies of this system posit that the analysis of short text messages and keyboard typing patterns can provide emotion information that may be used to support both clients and responders. 
We use our preliminary findings to discuss the way forward for applying AI to support mental health providers in providing better care.
\end{abstract}




\maketitle

\section{Introduction}
Mental health conditions and suicide are rising global health issues. 
In 2023, suicide was reported as the fourth leading cause of death for youths aged 15 to 19, with more than 700,000 people dying due to suicide every year~\cite{who-suicide-2023}. 
In response to the growing need for mental health services, online text-based helplines have emerged as a convenient and accessible option for individuals in distress, offering support around the clock~\cite{hoermann-2017-textbaseddialoguementalhealth}.
These platforms can be managed by AI-powered bots, licensed professionals, or trained staff. 

Our work aims to inform the design of messaging platforms for mental health conversations, especially those used by manned helplines, by understanding the emotions of its users. 
As studied by \cite{bhattacharjee-2023-textwellbeing}, different factors including the client's affective states must be considered when building context-aware text messaging systems for mental health. 
Additionally, human responders who find fulfilment in helping clients may also face distress and unintentionally affect their quality of care~\cite{willems-2020-crisisline}.
We investigate how incorporating emotion detection capabilities into these platforms could ease the burden on them and support more effective client communication.

Toward this goal, we are developing an unobtrusive emotion detection system, integrating both keystroke typing patterns, i.e., keystroke dynamics (KD), and sentiment analysis, to fit seamlessly into existing text-based support platforms. Both analyses can be obtained  without changing the inherent interaction mechanics of users.
We also aim to explore how the emotions detected can be interpreted and presented.

We are motivated by two main research questions: 
\begin{enumerate}
    \item Can KD and text data signal different emotions within a synchronous text conversation?
    \item What types of feedback can be presented to mental health helpline responders based on their own emotions and the client's?
\end{enumerate}
This paper presents the initial findings from (1) and its implications on AI-supported interventions in mental health and crisis care.
A within-subjects experiment was conducted with 31 participants who used a text messaging interface to engage in a synchronous conversation with a member of the research team serving as a moderator. 
Each participant's text and KD data were collected and analysed using machine learning for emotion detection.
\section{Related Work}
Previous studies have explored various methods and benefits of incorporating emotion capabilities in text messaging services. 
For example, \cite{nguyen-2022-moodyman} investigated the integration of emotional analysis tools into an online communication platform and found that making emotion information available within the chat can enhance teamwork and relationship building. 
\cite{wang-2004-physiologicalsensors}'s work with galvanic skin response sensors combined with user-selected tags indicates that visual feedback on a chat partner's emotions can make conversations more engaging.

KD, which initially garnered interest as a biometric form of user authentication, has been examined for its potential to reveal traits such as honesty, gender, and personality characteristics with varying degrees of accuracy. 
Notably, studies~\cite{borj-2019-kdliars, monaro-2017-kdliars} demonstrated that KD could predict whether individuals were lying, while \cite{li-2019-kdgender, buker-2019-kdgender, buker-2021-kdpersonality} utilised keystroke data combined with stylometric features to predict users' gender and personality traits. 
Similar to these demonstrations of the potential of KD as a rich source of behavioural biometrics, other research such as those by \cite{epp-2011-kdemo} has also shown KD as a promising method for detecting different emotional states.
Recent studies~\cite{maalej-2020-kdreview, yang-2021-emorecognitionreview} have built upon this, broadening the list of viable keystroke features and advanced analytical techniques.

Studies using KD for emotion detection commonly recommend using a multi-modal approach, i.e., fusing KD with other user data, for higher accuracy \cite{maalej-2020-kdreview}. 
Despite progress in combining KD with text analysis~\cite{nahin-2014-kdandtext, tahir-2022-kdandtextdatabase}, real-time application in synchronous chat contexts remains underexplored. 
This gap presents a unique opportunity for our work to contribute by applying these methodologies specifically to engaged conversations for mental health care.
\section{Methodology}
\subsection{Study Design}
We devised a study that emulates a text-based mental health support platform to evaluate the efficacy of passive emotion detection methods. 
We developed a chat messaging application typical of mental health support helplines and had participants interact with a moderator through the application in a desktop computer setting. 

Participants were asked to watch a short video intended to invoke a specific emotional reaction.
After viewing the video, they discussed their emotional responses with the moderator.
This activity was conducted three times per participant with three unique videos (randomly selected from our pool of eight), after which the moderator would bring the conversation to a close. 
The pool of eight short videos (42-62 seconds) was carefully chosen from the OpenLAV dataset~\cite{openlav-2021} to elicit different emotional responses spanning all combinations of valence (negative, neutral, positive) and arousal (low, medium, high).
This approach ensures a comprehensive range of emotional experiences for each participant and the overall study.

The chat application recorded all user key presses and text messages. 
Participants were assigned unique numbers to serve as identifiers within the study, maintaining user confidentiality. 
Any personally identifiable information (PII) used to conduct the experiments were stored separately from all study data and discarded afterwards. 

The study involved 31 participants (18 female), aged 18-50 (mean age = 28.7, SD = 8.3), with basic computer proficiency and who use English as their primary input language.
Since the emotion stimuli may trigger high levels of distress in participants, we excluded people with clinical mental health diagnoses like post-traumatic stress disorder or anxiety to minimise this risk in the initial study.
Participants were compensated with \$10 vouchers, and the protocol was approved by our university's Institutional Review Board (IRB).

\subsection{Data Analysis}
In classifying emotions, there are several models that either characterise them within a \textit{dimensional} space or differentiate emotions as discrete \textit{categories}.
These models provide us a theoretically grounded framework for understanding emotional states in mental health communications. 
For the \textit{dimensional} model, we applied the circumplex model of affect~\cite{posner-2005-circumplex} to classify the data points along a two-dimensional scale of valence and arousal, referring to the pleasantness of a stimulus and its intensity respectively.
For the \textit{categorical} model, we used Ekman's basic emotions~\cite{ekman-1999-emotions} plus a non-emotion state to classify our data into seven categories: surprise, happiness, sadness, anger, disgust, fear, and neutral. 
Both models, offering a clear lens through which to view the complex landscape of human emotions, have been used in emotion detection methods using KD~\cite{epp-2011-kdemo, maalej-2020-kdreview, nahin-2014-kdandtext}.

\subsubsection{Text Analysis} 
All our analyses rely on manual annotations as ground truth.
Two annotators from the research team labelled each message sent by the participants with \textit{dimensional} and \textit{categorical} emotion values.
Valence and arousal were both labelled on a scale of -1, 0, and 1 to denote negative to positive emotional states and low to high intensity levels, respectively.
Up to three emotion labels were assigned per message for the \textit{categorical} classification of emotions.
To ensure the reliability of the manual annotations, we computed the Krippendorf's $\alpha$ as a measure of inter-annotator agreement. We obtained a score of 0.75 for valence, 0.56 for arousal, and an average of 0.76 across the seven emotion labels. Discrepancies during the aggregation of the annotations were resolved by a random selection of labels from either annotator.

Utilising OpenAI's GPT-4 Turbo (gpt-4-0125-preview), we harnessed the strengths of Large Language Models (LLMs) in nuanced emotion detection and linguistic adaptability by providing the same dataset and instructions as the human annotators, i.e., the full conversation logs with messages from the participants and the moderator. 
In this approach, the GPT-4 LLM also generated \textit{valence} and \textit{arousal} values and up to three dominant emotion labels. 
These were used for predicting the manually annotated emotional states.

\subsubsection{Keystroke Dynamics}
We transformed raw keystroke data from the in-person experiments into a feature set that could be used as input to predictive models, applying a method similar to \cite{epp-2011-kdemo}.
We extracted two categories of features: keystroke and content features. 

Keystroke features included statistical measures of timing-related attributes such as intervals between key presses and releases, nuanced typing habits such as \emph{backspace frequency} to infer correction habits and \emph{Enter} key usage to understand message pacing. 

Content features encompassed punctuation, capitalisation, and sentence structure to observe authentic conversational typing habits in contrast to controlled text inputs. 
These features, combining both mechanical and compositional typing aspects, highlight the user's natural interaction patterns and are pivotal in building machine learning models that can predict different emotional states. 

Among random forest (RF) classifiers, support vector machines (SVMs), and multinomial logistic regression models, we achieved the best results using RF (average precision=0.796, F1 score=0.791).

\subsubsection{Fusion of Text Analysis and KD} 
We combined the results from the LLM-based text analysis with KD to yield features which were used in another set of machine learning models that predicted the same labels. 
This fusion of features, specifically taken in a conversational context, is the novel method of emotion detection we are presenting in this work.
\section{Preliminary Findings}
In this study, we evaluated the performance of machine learning models that predicted \textit{valence} and \textit{arousal} levels and emotion categories in terms of overall accuracy, precision, recall, and F1 score. 
We compared the results between models that use features from the LLM-based text analysis (henceforth referred to as text-only models), models that use only KD features (henceforth referred to as KD-only models), and models that use a fusion of both (henceforth referred to as fusion models). 

From our interpretation of the results (partially presented in Table \ref{tab:results-summary}), models using the \textit{categorical} emotion classification achieved better results than those using the \textit{dimensional} one in general. 
Even the lowest performance from a \textit{categorical} model (for \textit{neutral} emotion, precision=0.756, F1 score=0.756) was higher than the better performing \textit{dimensional} model (for \textit{arousal}, precision=0.746, F1 score=0.746).
In predicting \textit{happiness}, \textit{surprise}, \textit{fear} and \textit{anger}, and the \textit{valence} level, the fusion models had higher scores than any of the standalone counterparts.
However, we noted that in these cases, the differences between the text-only and fusion models are small (mean $\Delta$F1 score=0.007), revealing that the contribution of KD could be minimal.
Additionally, for the rest of the emotion classifications, text-only models either performed better than the fusion models (i.e., for \textit{neutral}, \textit{sadness}, \textit{disgust}) or similarly (i.e., for \textit{arousal}). 
Overall, the best results were from the fusion models predicting \textit{anger} (precision=0.984, F1 score=0.985), \textit{surprise} (precision=0.976, F1 score=0.967), and \textit{fear} (precision=0.956, F1 score=0.956).

\begin{table*}[t]
    \caption{Summary of results from the fusion models}
        \label{tab:results-summary}
    \begin{tabular}{cccccccccc}
        \toprule
        \textbf{} & \textbf{Valence} & \textbf{Arousal} & \textbf{Neutral} & \textbf{Happiness} & \textbf{Sadness} & \textbf{Disgust} & \textbf{Fear} & \textbf{Surprise} & \textbf{Anger} \\ 
        \midrule
        \textbf{Accuracy} & 0.660 & 0.791 & 0.756 & 0.830 & 0.938 & 0.956 & 0.961 & 0.975 & 0.988 \\ 
        \textbf{Precision} & 0.660 & 0.746 & 0.756 & 0.831 & 0.933 & 0.951 & 0.956 & 0.976 & 0.984 \\ 
        \textbf{Recall} & 0.660 & 0.791 & 0.756 & 0.830 & 0.938 & 0.956 & 0.961 & 0.975 & 0.988 \\ 
        \textbf{F1 score} & 0.660 & 0.746 & 0.756 & 0.830 & 0.931 & 0.950 & 0.955 & 0.967 & 0.985 \\ 
        \bottomrule
    \end{tabular}
\end{table*}

We integrated one of the trained KD-only models with our instant messaging application to verify that it can be utilised in an online chat context.
The model was first converted into a format suitable for real-time inference and prediction of emotion values in our prototype. 
After a message is sent, the model successfully predicts \textit{valence} and \textit{arousal} values using keystroke features from the user.
This capability can be further developed with LLM integration to perform emotion analysis using the fusion model. 
While additional testing needs to be done to enhance predictive accuracy, our early results demonstrate a promising outlook in conveying emotional information instantaneously in engaged mental health conversations. 
\section{Discussion}
\subsection{Improving Predictive Models}
The results from the predictive models using KD indicate that there is room to increase accuracy given some modifications to our methodology. 
For example, the data can be analysed at a different level of granularity, i.e., per-chunk of consecutive messages or per-time-based intervals rather than per-message. 
Other keystroke features such as those based on digraphs and trigraphs commonly found in the local language \cite{li-2019-kdgender, epp-2011-kdemo} can also be extracted for analysis.
More importantly, aligned with the rapid evolution of AI and the growing amount of high-performing algorithms, other feature selection techniques and machine learning classifiers can be explored. 
As shown in other studies, deep learning methods can be used to analyse KD for emotion recognition~\cite{yang-2021-emorecognitionreview}.

\subsection{Comparison and Integration with Other Physiological Measures}
Our work focused on emotion detection through text data and KD, yet the integration of physiological measures could deepen our understanding of emotional responses. 
During our study, we collected facial data via webcam and heart rate data via smartwatch, as studies have shown that these can be used to detect emotional states~\cite{canedo-2019-fer} and measure stress levels~\cite{hickey-2021-smartdevicesmentalhealth}, respectively. This additional data and multi-modal approach could enrich our comparison of text and KD analysis with existing methods of understanding affective states.

Although our primary goal is unobtrusive emotion detection, leveraging increasingly ubiquitous devices such as smartwatches might enhance the accuracy of our proposed models.
Future research could explore how additional physiological data alters our findings or provides new insights into how emotions affect communication patterns.

\subsection{Understanding the Impact of Emotions in Engaged Mental Health Conversations}
\begin{figure*}[t]
    \centering
    \includegraphics[width=0.8\textwidth]{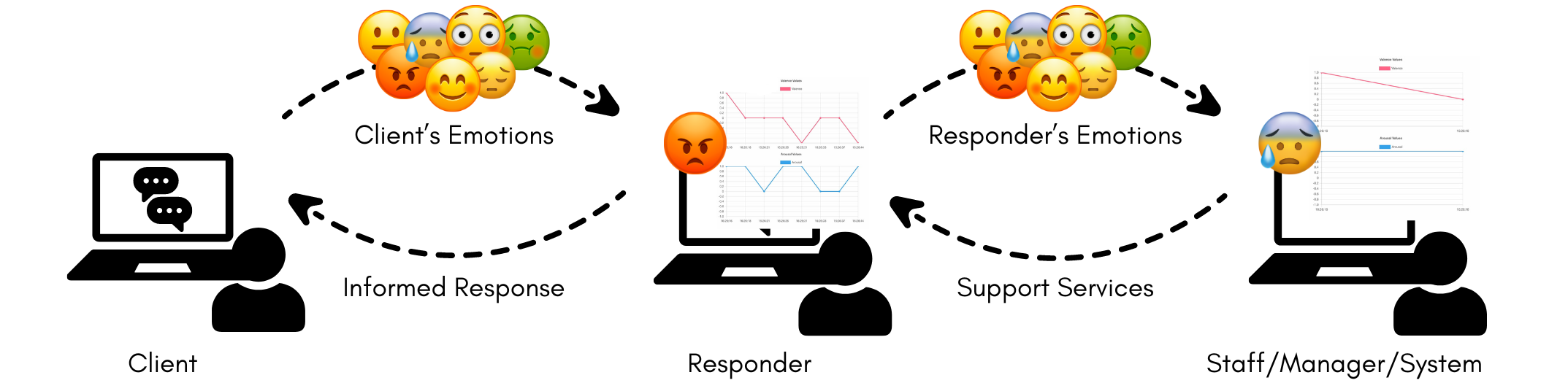}
    \Description[Relationship(s) between users on an Emotion-Aware Platform]{Conceptual Diagram Showing the Relationship(s) between a Client, Responder and Staff/Manager/System on an Emotion-Aware Platform}
    \caption{Concept of an emotion-aware platform for mental health conversations \copyright Kellie Yu Hui Sim; Kohleen Tijing Fortuno; Kenny Tsu Wei Choo.}
    \label{fig:platform-concept}    
\end{figure*}

Ultimately, we aim to propose design considerations for emotion-aware online platforms (Figure \ref{fig:platform-concept}), on which engaged mental health conversations can take place. 
Providing responders with this dynamic tool to accurately assess clients' emotional states in real-time may have several benefits, including guided decision-making for urgent care or intervention and an enhanced quality of support. 
Such platforms can also detect distress in responders and offer response suggestions or self-care advice where necessary. 

It is important to note that this technology is designed to aid human-supported services rather than replace the capabilities of a human crisis helpline responder. 
As with other AI-generated content, insights provided by this emotion detection system must be judged with vigilance and without forgetting the value of human empathy.
We recommend that the system design implements features for offering suggestions rather than asserting facts, such as presenting confidence scores alongside predictions.

Future research could explore the effectiveness and evaluate the risks of various methods in which emotion-related information and other relevant insights are presented to users, and closely collaborate with mental health professionals to ensure our model's seamless integration into existing frameworks. 

\subsection{Limitations and Challenges}
We excluded modern text input capabilities such as auto-correct, auto-complete, and emojis to keep a refined approach for our study.
Minimising other methods of text input allowed us to capture keystrokes that closely match with the recorded text messages, simplifying our data pre-processing steps.
As this may have impacted participants' usual flow of expression, future research might incorporate these input methods to assess their influence on KD and the overall message sentiment.

Furthermore, the limited data from a controlled setting hinders the creation of personalised models that account for a user's unique texting and typing habits.
Such tailored models can be more sensitive and responsive to subtle shifts in emotional states and can be achieved through a more extensive user study.

The use of LLMs to analyse text messages which may contain sensitive information raises data privacy and consent issues. 
Similar to the steps we took to protect participants in our study, we propose the inclusion of confidentiality measures such as data obfuscation, encryption, and secure storage. 
As observed in our study, user personal information is not necessary to predict emotions and may be removed or obfuscated in text messages before being provided to the models for processing. 
Further research could explore fine-tuning an open-source LLM for localised mental health services and running the system in private servers only accessible to the crisis care provider. 
To add an extra layer of protection, the system should also encrypt all data at rest and in transit.
Running the analysis in real-time also means that raw data will be immediately deleted and only the detected emotions, free of any potentially sensitive information, are stored. 
\section{Conclusion}
In this paper, we uncovered the benefits of an emotion-aware messaging platform in supporting engaged mental health conversations.
We investigated the use of textual data and keyboard typing patterns as an unobtrusive method of detecting emotions under \textit{dimensional} and \textit{categorical} classifications.
While our work showed promise in recognising emotion categories, especially with the help of LLMs for sentiment analysis, there is room to improve the contribution of KD in our predictive models.
We conclude that emotion detection capabilities can be feasibly integrated into existing messaging platforms, yet the applications of this technology must be designed with care, especially in mental health and crisis response settings.
Our work not only enhances technical methodologies in emotion detection by leveraging AI but also paves the way for more empathetic and responsive mental health services, setting a new standard for future digital communication and care.

\bibliographystyle{ACM-Reference-Format}
\bibliography{references}

\end{document}